# Anomalous Ultrasonic Attenuation in Aqueous NaCl Solutions

Barnana Pal and Srinanda Kundu
Saha Institute of Nuclear Physics, 1/AF Bidhannagar,
Kolkata –700064, India.

**Abstract**:

The velocity (v) and attenuation constant ($\alpha$) for ultrasonic waves of frequencies 1MHz and 2MHz propagating through aqueous sodium chloride solution have been measured over the concentration (c) region 0–5.3 $mol.L^{-1}$ at room temperature (25$^0$ C). The velocity (v) shows an overall increase with the increase of c indicating comparatively stronger bonding among the ions and water molecules prevailing in the solution. The attenuation constant, besides showing an overall increase with c, shows significantly high values at some concentrations. Attempt has been made to understand the behaviour from existing theoretical background.

## 1. Introduction

The study of the properties of aqueous electrolytes with different concentrations varying from very low to supersaturated conditions has got importance primarily due to their complex character. An elaborate study of the physical and chemical properties of these systems is important not only for their significant technological and industrial applications, but also for the basic understanding of their nature from a microscopic point of view. Till date lot of theoretical and experimental works have been done on the dynamical and structural properties of such systems [1-8]. The experimental techniques used for such studies include Laser Raman Spectroscopy [8,9], Dynamic Light Scattering (DLS)[10], Neutron Diffraction [11], Ultrasonic Spectroscopy [12,13], and so on. Still, the behaviour of such systems with different degree of dilution remains unexplored mainly due to the lack of information regarding the interaction potential between the ions and water molecules present in the solution.

It is known that water is a strong polar liquid with dipole moment 1.86D and in pure water the water molecules rearrange themselves in hydration shells [14]. However, the spatial extent of such shell is not well known. Also for electrolyte solutions, there are no well-defined structural model or well-substantiated parameters for ion hydration spheres [12]. There are evidences of cluster formation in such electrolytes. Laser Raman Spectroscopic studies in unsaturated, saturated and supersaturated solutions indicate the presence of clusters of different sizes [8,9]. As the solution concentration increases, rearrangement between the prevailing ions and clusters takes place due to coalescence of some clusters. Direct evidence for the formation of large size clusters is obtained from DLS experiments [10], where clusters having size distributions with hydrodynamic radii below 1nm and in between 300 to 500nm have been detected. The X-ray and neutron diffraction experiments [11] are unable to detect large size clusters but the autocorrelation functions obtained from these experiments are clear indications for the formation of large size clusters.

Computer simulation of these systems using Molecular Dynamics (MD) method [15-21] and Monte-Carlo (MC) technique [22-24], however, seems to be useful in understanding the system from a microscopic point of view. MD study [20] clearly indicates that the probability of formation of large size clusters increase with the increase in concentration of the solution.

NaCl is one very common but important electrolyte, which dissociates into $Na^+$ and $Cl^-$ ions in water. Sound absorption in aqueous electrolytes including NaCl has been studied [25] over a wide frequency range from 10 kHz to 300 MHz in the low concentration region (0.01 mol·$L^{-1}$ to 0.1 mol·$L^{-1}$) where it is found to be independent of solution concentration. The present report deals with the ultrasonic study on aqueous NaCl solution at higher concentrations ranging from 0 to 5.3 mol·$L^{-1}$. The variations of the velocity of propagation (v) and that of the attenuation constant (α) have been measured at room temperature (~25$^0$C). Section 2 presents a brief description of the method. Results and discussion are presented in Section 3 and Section 4 respectively and conclusion is given in Section 5.

## 2. Experimental method:

Measurements of v and α of ultrasonic waves with frequencies 1 MHz and 2 MHz have been done using standard pulse-echo method [26]. High frequency rf bursts with carrier frequencies 1 MHz and 2MHz have been generated using ULTRAN HE 900 pulser-receiver. The pulse width is ~5μs with a repetition time ~10ms. Peak-peak height of the pulses are ~300V. NaCl solutions of concentrations ranging from 0.16 mol·$L^{-1}$ to 5.31 mol·$L^{-1}$ have been prepared using 99.9% pure NaCl in millipore deionized water. The estimated error in the solution concentration (c) is less than 1%. The solution is taken in a cylindrical glass container with optically flat inner bottom. The mounted quartz transducer placed at the top of the container acts as the transmitter as well as the receiver for the ultrasonic wave propagating through the solution. The echoes are captured on a scopemeter (Model FLUKE 199C) and from the positions and heights of the echoes, the velocity and attenuation are calculated using standard curve fitting method.

Measurements at each frequency have been done three to seven times at each concentration with freshly prepared solutions as well as sonicated solution. Sonication has been done at 40 kHz for 30 minutes in a sealed flask to remove air bubbles present in the solution. At least five observations are made at each run. We observe that the velocity values obtained for different sets of experiments are nearly equal but attenuation values fluctuate. At the concentrations where fluctuation is more the experiment is repeated five to seven times, with at least five observations. Experiments have been done at room temperature 25ºC±0.1ºC.

## 3. Experimental Results:

The variation of v with c is presented in figure1. Fig 1(a) gives the result for 1MHz wave and 1(b) is that for 2MHz wave frequency. Velocity values multiplied by $10^{-5}$ are given in cm.sec$^{-1}$ and c in mol.L$^{-1}$. For both the frequencies v increases with the increase of c. The variation of v with c fits very well to a relation given by,

$$v = A_0 + A_1.c + A_2.c^2 \qquad (3.1)$$

with $A_0 = 1.496 \pm 0.003$, $A_1 = 0.069 \pm 0.002$ and $A_3 = -0.0028 \pm 0.0004$.

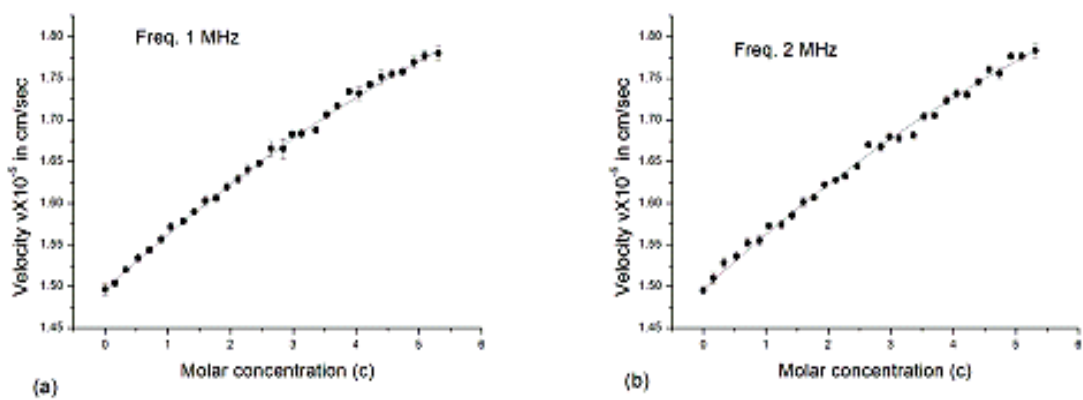

*Figure-1: Plot of velocity (v) multiplied by $10^{-5}$ in cm.sec$^{-1}$ as function of concentration(c) in mol.L$^{-1}$. (a) is for ultrasound with frequency 1 MHz and (b) is for 2MHz ultrasound frequency.*

Figure 2 depicts the variation of $\alpha$ in db.µs$^{-1}$ with c for (a) 1 MHz and (b) 2 MHz waves. The attenuation constant, in addition to an overall increase with the increase of c, rises significantly at concentrations 2.46 M.L$^{-1}$ and 3.89 M.L$^{-1}$, though the rise at 2.46 M.L$^{-1}$ is not too prominent for 1 MHz wave. If the existence of the attenuation peaks is ignored, the dependence of $\alpha$ on c can be represented as,

$$\alpha = B_0 + B_1.c^2 + B_2.c^3 \qquad (3.2)$$

The parameter $B_0 = 0.100 \pm 0.004$ for 1 MHz and $0.088 \pm 0.004$ for 2 MHz. $B_1$ and $B_2$ are same for the two frequencies with values $0.004 \pm 0.0002$ and $-0.0005 \pm 0.00004$ respectively.

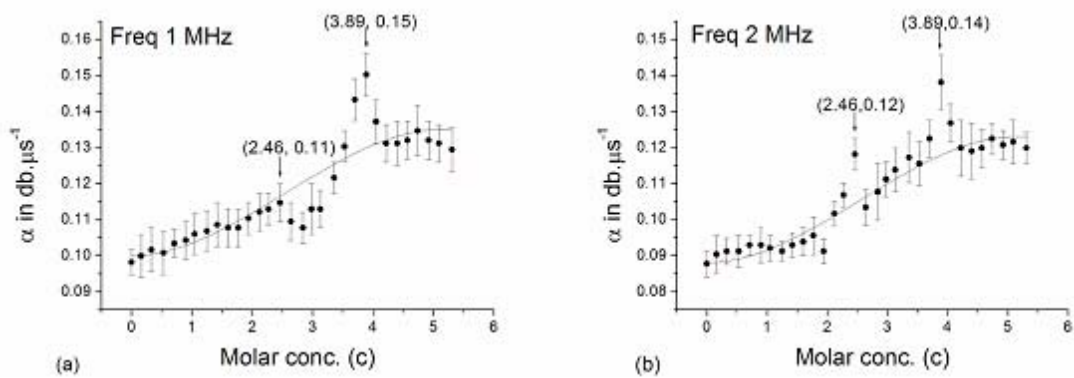

*Figure-2: Plot of attenuation constant (α) in db·μs$^{-1}$ as function of concentration (c) in mol·L$^{-1}$ for ultrasound frequency 1MHz. 2(a) presents data points obtained for three different sets of experiments and (b) gives the average over all observations.*

Measurements with sonicated solutions show no significant change in velocity values for both the frequencies. Also α remains unaffected at 1 MHz. For 2 MHz wave α remains almost unaffected for all concentrations except near the peak positions, where we found α to reduce a little, the reduction being less than 10%.

**4. Discussion**:

Figure 1 shows that v increases with the increase of c both for 1MHz and 2MHz waves, indicating a stronger bonding among the prevailing ions and water molecules in the solution. The velocity values are in good agreement with the measurement by Gucker et. al. at 4 MHz[13]. The attenuation plots show the presence of two distinct attenuation peaks both for 1MHz and 2MHz waves. Table-1 gives the concentration values along with the corresponding velocity and wavelength values at which these peaks are located.

***Table-I***: *Solution concentrations and corresponding velocity and wavelength values at which attenuation peaks are observed for 1MHz and 2MHz ultrasonic wave frequencies.*

| Attenuation peak no. | Conc. c (mol·L$^{-1}$) | Observation for ultrasonic wave with frequencies | | | |
|---|---|---|---|---|---|
| | | 1 MHz | | 2 MHz | |
| | | Velocity v×10$^{-5}$ (cm.sec$^{-1}$) | Wavelength λ (cm) | Velocity v×10$^{-5}$ (cm.sec$^{-1}$) | Wavelength λ (cm) |
| 1 | 2.46 | 1.6477 | 0.1648 | 1.6442 | 0.0822 |
| 2 | 3.89 | 1.7339 | 0.1734 | 1.7235 | 0.0861 |

The mechanisms responsible for ultrasonic attenuation in liquids are, (i) viscous loss, (ii) loss due to heat conduction and (iii) scattering loss. The general theory that takes into account the effect of viscosity and thermal conduction gives [27],

$$\alpha = (\omega^2/2\rho v^3)[(4\eta/3+\eta^v) + \tau (1/C_p - 1/C_v)] \tag{4.1}$$

where $\omega$ is angular frequency, $\rho$ is density, $\eta$ is shear viscosity, $\eta^v$ is volume viscosity, $\tau$ is thermal conductivity, $C_p$ and $C_v$ are specific heats at constant pressure and at constant volume respectively. We observe that with the increase of c, $\rho$ increases linearly and v increases showing nonlinear behaviour at higher c values. $\eta$ and $\eta^v$ increase exponentially up to c ~ 3 M.L$^{-1}$ and at higher solution concentrations the rate of increase is lower [28,29]. The empirical relation, viz., the Arrhenius-Andrade relation, $\eta = A \exp(E/RT)$, where A and E are concentration dependent but temperature independent parameters and T is the temperature, gives the viscosity variation at lower c. $\tau$ decreases with the increase of c [2,3] and the difference is ~7% of 5 M.L$^{-1}$ NaCl concentration. $C_p$ is a little higher than $C_v$ for NaCl solution and both of them decrease with the increase of c [30]. Thus, from eqn. (4.1) it is clear that the dominating effect causing variation of $\alpha$ with c comes from the variation of $\eta$ and an exponential increase of $\eta$ with c as given by Arrhenius-Andrade relation may give rise to an overall increase of $\alpha$ with c. However, the viscous loss and thermal effect is insufficient to explain the significant increase in $\alpha$ at particular c values observed in the present experiment.

Another process we have not considered yet is the scattering phenomena. In the present experiment we use pulses with carrier frequencies 1MHz and 2MHz. A carrier frequency repeating pulse actually is composed of an infinite number of component waves with frequencies and amplitudes given by the Fourier components of the pulse signal. Among them only those components contained within the bandwidth of the transducer and its associated electrical circuit will contribute to the formation of echoes [31]. Usually the bandwidth is ~10% of the carrier frequency. If we consider this then for 1MHz signal, waves with wavelength ranging from 0.136cm to 0.166cm in pure water and from 0.162cm to 0.198cm in saturated NaCl solution will be present in the bandwidth. For 2 MHz pulse signal, these are from 0.071cm to 0.079cm in pure water and 0.085cm to 0.094cm in saturated NaCl solution. Thus scattering of some of these component waves by the loosely bound fractal like ion-solute clusters formed in the solution may be responsible for the observed attenuation rise. Theoretically it has been shown that the attenuation due to scattering from non-rigid spherical particles with radius 'a' suspended in a medium is given by [ref 27, p-143],

$$\alpha = \phi\omega^4 a^3/(2v_m^3)[1/3\{1 - \rho_m v_m^2/(\rho_p v_p^2)\}^2 + \{(\rho_p - \rho_m)/(2\rho_p + \rho_m)\}^2 \tag{4.2}$$

where $\phi$ is the volume fraction of scattering particles. Subscripts p and m stand for the particle and the medium respectively. Eq. (4.2) is valid in the long-wavelength limit ka<<1. Also it has been shown that the scattering cross-section is significant only for ka>0.2 [ref 27, p-139]. In the present experiment, the ultrasound wavelengths corresponding to the attenuation peaks lie in the range 0.0822 cm to 0.1734 cm (Table-1) and if it is assumed that scattering due to ion-solute clusters are responsible for the observed attenuation peaks, then the effective size of such clusters will be ~25 micron for 2MHz wave and ~50 micron for 1MHz wave. This is nearly 100 times larger than those detected in DLS experiments [10] that provide direct evidence for the existence of clusters of distinct well-defined sizes. In this experiment $Ar^+$ laser at wavelength 488 nm has been used to study NaCl solution with different concentrations ranging from 0.68 $mol \cdot L^{-1}$ to 5.32 $mol \cdot L^{-1}$ (saturated). Particles having size distributions with hydrodynamic radii below 1nm and in between 300 to 500nm have been detected. Particles having larger dimensions, if there are any, may not be detectable in this experiment because of the limitation due to the source wavelength.

There are other experimental results providing evidences for the existence of large size clusters in aqueous electrolytes [8-11] and even in pure water [14]. The partial pair correlation functions (ppcf) obtained in Neutron diffraction experiments [11] clearly indicate the formation of water clusters with definite order. The size of such clusters, however, cannot be predicted from these experiments but, definitely, it can be inferred that large size clusters are to be present in the system to produce the experimentally observed ppcf. Laser Raman Spectroscopic studies in unsaturated, saturated and supersaturated solutions indicate the presence of large clusters of different sizes [8,9]. Study with aqueous $NaNO_3$ solution [9] shows that such clusters do not have crystalline structure. These are clusters consisting of solute and solvent molecules. With the increase of solution concentration, rearrangement between the prevailing ions and clusters takes place and due to coalescence of some clusters, larger clusters are formed with consequent dilution of other clusters. Once a large size cluster is formed it becomes a stable one. These studies, however, cannot predict the order of magnitude of the size of these clusters. But, the wavelength of the vibrational mode detectable in Laser Raman Spectroscopic studies [9] is found to be ~ 0.006 cm. Also the Far Infra-Red (FIR) spectroscopic study in pure water [14] gives a translational mode with wavelength as large as 0.03 cm. These are indications for the existence of larger size clusters in the electrolyte solution as well as in pure water. The neutron diffraction experiments [11] cannot detect large size clusters. But the autocorrelation functions obtained in these experiments provide indirect evidence for the formation of large size clusters. Further, measurement of diffusion constant of Na in aqueous NaCl solution [32] shows the diffusion constant to decrease with the increase in solution concentration and the decrease is more rapid in the higher concentration ($c > 1$ $mol \cdot L^{-1}$) region. Formation of large size ion-water clusters in this concentration region may be a suitable explanation for the decrease in diffusion data.

Formation of large size clusters is also supported by simulation studies using Molecular Dynamics method [19,20] where it has been shown that clusters with more than twenty ions can be formed and the probability of formation of large clusters increase with the increase of solution concentration. Thus our observation indicating the possibility of formation of large size clusters at these particular concentrations is quiet justified. Obviously, clusters of different sizes will be formed in the solution, i.e. a size distribution of clusters will be observed.

Another observation of particular interest is the formation of bubbles in the solution as a consequence of acoustic cavitation [23]. Due to the propagation of ultrasonic waves bubbles are formed inside the solution and they stick to the inside walls of the container. Air molecules are always present to some extent in the electrolytic solutions. During the propagation of ultrasound pressure fluctuates from a minimum to a maximum value in one half cycle. Near the minimum pressure region the air expands and the nearby air bubbles coalesce to form bubbles of larger size. Some of these bubbles may have such a size as to execute resonant oscillation at the imposed frequency, giving rise to stable cavitation and a consequent increase in the attenuation constant. To study the effect of dissolved air bubbles, we repeated the experiment with the solutions sonicated at 40kHz for 30 minutes. Only a little change has been detected for the attenuation values near the peak positions at 2MHz. We noticed a reduction in the attenuation values, the reduction being less than 10%. The size of the air bubbles present in the solution has been calculated considering surface tension of pure water–air interface and saturated NaCl solution–air interface. The radii of largest bubbles in pure water and saturated NaCl solutions under present experimental conditions are 0.0367 cm and 0.0342 cm respectively and these may contribute a little towards attenuation for 2MHz wave. However, for a better understanding of the observed phenomenon an extensive study of the velocity and attenuation properties with ultrasound in the frequency range from KHz to MHz region is necessary. Also study of the temperature variation of the phenomena is important in this issue. Rise in temperature will cause the larger clusters to break up due to thermal agitation leading to a consequent decrease in the peak heights.

**5. Conclusion:**

The present study on the concentration dependence of velocity and attenuation of ultrasonic waves with carrier frequencies 1MHz and 2MHz pulsed signal indicate a stronger bonding among prevailing ions and solvent molecules with increasing solute concentrations. Significant rise in the attenuation value may be due to the formation of ion-solvent clusters in the unsaturated and saturated solutions. Air bubbles may be present in the solution. The estimated largest radius of such bubbles is 0.0367 cm in pure water and 0.0342 cm in saturated solution. These bubbles may contribute to the attenuation at 2 MHz. Till now there is no direct experimental evidence for the formation of clusters having such a large dimension. The only direct evidence is obtained from DLS experiment where the estimated size is below 1 nm and also between 300 to 500 nm. Raman Scattering experiments, Neutron Diffraction experiments and FIR spectroscopic studies provide strong indications for the formation of large size clusters in aqueous electrolytes as well as in pure water. These experiments, though cannot provide any estimate regarding the actual size of the clusters, clearly point towards the formation of very large size clusters, even of the order of mm. However, more study regarding the temperature variation of the phenomena and study with other ultrasound frequencies are necessary to gain better understanding of the present observation.

**Acknowledgement**:

The authors are grateful to Dr. Haimanti Chakrabarti for suggesting the problem and to Mr. Anis Karmahapatro for related technical assistance.

**References:**

[1] R. A. Robinson, R. H. Stokes, Electrolyte Solutions: The Measurement and Interpretation of Conductance, Chemical Potential and Diffusion in Solutions of Simple Electrolytes, Academic, New York, 1959.
[2] M L V Ramires, C A Nieto de Castro, J M N A Fareleira and W A Wakeham, Thermal conductivity of aqueous sodium chloride solutions. J. Chem. Eng. Data **39,** 186-190 (1994).
[3] V D Yusufova, R I Pepinov, V A Nikolaev and G M Guseinov, Thermal conductivity of aqueous solutions of NaCl. J Eng Phys & Thermophys. **29**, 1225-1229 (1975).
[4] J. Barthel, R. Buchner, Dielectric properties of nonaqueous electrolyte solutions. Pure & Appl. Chem., **58**, (1986) 1077-1090.
[5] J. Barthel, R. Buchner, K. Bachhuber, H. Hetzenauer, M. Kleebauer, H. Ortmeier, Molecular processes in electrolyte solutions at microwave frequencies. Pure & Appl. Chem., **62,** (1990) 2287-2296.
[6] J. Barthel, R. Buchner, High frequency permittivity and its use in the investigation of solution properties. Pure Appl. Chem., **63,** (1991) 1473-1482.
[7] R. Buchner, J. Barthel, Dielectric relaxation in solutions. Annu. Rep. Prog. Chem., Sect.C, **91,** (1994) 71-106.
[8] M. K. Cerreta, K. A. Berglund, The structure of aqueous solutions of some dihydrogen orthophosphates by laser Raman spectroscopy. J. Cryst. Growth, **84,** (1987).
[9] Rusli, I. T., Schrader, G. L., Larson, M. A.: Raman spectroscopic study of NaNO$_3$ solution system – solute clustering in supersaturated solutions. J. Cryst. Growth, **97,** 345-351 (1989) 577-588.
[10] Y. Georgalis, A. M. Kierzek, W. Saenger, Cluster formation in aqueous electrolyte solutions observed by dynamic light scattering. *J. Phys. Chem*., **104**, (2000) 3405-3406.
[11] S. E. Mclain, S. Imberti, A. K. Soper, A. Botti, F. Bruni, M. A. Ricci, Structure of 2 molar NaOH in aqueous solution from neutron diffraction and empirical potential structure refinement. Phys. Rev. B, **74**, (2006) 094201-1-094201-7.
[12] B. Voleišienė, A. Voleišis, Ultrasound velocity measurements in liquid media. ISSN 1392-2114 Ultragarsas(Ultrasound), **63**, (2008) 7-19.
[13] F. T. Gucker, C. L. Chernick, P. Roy-Chowdhury, A frequency-modulated ultrasonic interferometer: Adiabatic compressibility of aqueous solutions of NaCl and KCl at 25°C, PNAS **55**, (1966) 12-19.
[14] F. N. Keutsch, R. J. Saykally, Water clusters: Untangling the mysteries of the liquid, one molecule at a time. PNAS, **98,** (2001) 10533-10540.
[15] L. Degreve, F. L. da. Silva, Large ionic clusters in concentrated aqueous NaCl solution. J. Phys. Chem., **111**, (1999) 5150-5156.
[16] D. M. Sherman, M. D. Collings, Ion association in concentrated NaCl brines from ambient to supercritical conditions: results from classical molecular dynamics simulations. Geochem. Trans., **3**, (2002) 102-107.
[17] D. Zahn, Atomistic Mechanism 102-107 of NaCl Nucleation from an Aqueous Solution. Phys. Rev. Lett., **92**, (2004) 040801-040804.
[18] D. Du, J. C. Rasaiah, J. D. Miller, Structural and Dynamic Properties of Concentrated Alkali Halide Solutions: A Molecular Dynamics Simulation Study. J. Phys. Chem. B, **111**, (2007) 209-217.

[19] A. A. Chen, R. V. Pappu, Quantitative Characterization of Ion Pairing and Cluster Formation in Strong 1:1 Electrolytes. J. Phys. Chem. B, **111**, (2007) 6469-6478.
[20] S. A. Hassan,: Morphology of ion clusters in aqueous electrolytes. Phys. Rev. E, **77**, (2008) 031501-1-031501-5.
[21] J. J. Molina, J-F. Dufrêche, M. Salanne, O. Bernard, M. Jardat, P. Turq, Models of electrolyte solutions from molecular descriptions: The example of NaCl solutions. Phys. Rev. E, **80**, (2009) 065103(R) 1-4.
[22] H. Chakrabarti, B. Pal, Diffusion in complex liquid surfaces – a Monte Carlo study. Indian J. Phys. A, **70**, (1996) 729-734.
[23] H. Chakrabarti, B. Pal, Diffusion in a strongly correlated 2D liquid system. Indian J. Phys. A, **78**, (2004) 935-937.
[24] H. Chakrabarti, B. Pal, Signature of glass transition in a strongly correlated 2D liquid. J. Phys: Condens. Matter, **18,** (2006) 9323-9334.
[25] K. Tamm, G. Kurtze, R. Kaiser, Measurements of sound absorption in aqueous solutions of electrolytes. Acoustica, **4**, (1954) 380.
[26] R. Truell, C. Elbaum, B. B. Chick, Ultrasonic Methods in Solid State Physics, Academic Press, New York & London, (1969) p-53.
[27] A S Dukhin and P J Goetz, "Ultrasound for characterizing colloids", Elsevier, Amsterdam, 2002, p-81.
[28] J. Kestin and I. R. Shankland, Viscosity of aqueous NaCl solutions in the temperature range 25–200 °C and in the pressure range 0.1–30 Mpa. Int J Thermophys. 5, 241-263 (1984)
[29] R. H. Stokes and R. Mills, Viscosity of electrolytes and related properties, International Encyclopedia of Physical Chemistry and Chemical Physics, **3,** (1965).
[30] K G Liphard, A Jost and G M Schneider, Determination of the specific heat capacities of aqueous sodium chloride solutions at high pressure with the temperature jump technique. J Phys Chem, **81,** 547-550 (1977).
[31] Barnana Pal, Response of a composite resonator under pulse excitation－a numerical study. Japanese Journal of Applied Physics Pt-1, **35** (1996) 4839.
[32] H. Chakraborty, Strong evidence of an isotope effect in the diffusion of a NaCl and CsCl solution. Phys. Rev. B, **51,** (1995) 12809-12812.
[33] T. G. Leighton, Bubble population phenomena in acoustic cavitation, Ultrasonics Sonochemistry, **2,** (1995) S123-S136 and references therein.